\documentclass{aa}
\usepackage[english]{babel}
\usepackage{latexsym,amsfonts,amssymb}
\usepackage{graphicx}

\usepackage{lscape}
\makeatletter
\def\fnum@figure{\small{\bf{Figure \thefigure}}}
\def\fnum@table{\small{\bf{Table \thetable}}}
\makeatother
\usepackage{indentfirst}

\def\mincir{\ \raise -2.truept\hbox{\rlap{\hbox{$\sim$}}\raise5.truept  
\hbox{$<$}\ }}                        

\begin{document}
\title{Radio Spectra of a Sample of X-ray Selected BL Lacs}

\author{Francesca Cavallotti\inst{1}, Anna Wolter\inst{1}, John T. Stocke\inst{2}, Travis Rector\inst{3} }  
\offprints{Anna Wolter, {\it anna@brera.mi.astro.it} }
\institute{Osservatorio Astronomico di Brera, via Brera 28, 20121, Milano, Italy, e-mail cavallot@brera.mi.astro.it \and Center for Astrophysics and Space Astronomy, University of Colorado, Boulder, CO 80309-0389, USA \and National Radio Astronomy Observatory, P.O. Box O, Socorro, NM 87801, USA; {\it present address}:
University of Alaska, 3211 Providence Dr., BMB 212, Anchorage, AK 99508, USA }

\date{Received  / Accepted }

\authorrunning{Cavallotti, F., Wolter, A. et al.}
\titlerunning{VLA spectra of EMSS BL Lacs}

\abstract{
We present simultaneous multifrequency radio observations
for a complete subsample of 26 XBLs
from the {\it Einstein} Extended Medium-Sensitivity Survey, obtained
with the Very Large Array (VLA).
Spectra are computed using fluxes at 20, 6 and 3.6 cm.
Unlike many radio selected samples, the EMSS did not impose any 
criterion on the radio spectrum to identify BL Lac objects.
It is therefore possible to investigate the intrinsic radio spectral slope 
distribution and to determine the effect produced by this
selection criterion. We find that 15\% of the observed objects do not meet the
flat-spectrum criterion imposed on some other BL Lac samples. 
A dataset that includes non-simultaneous
data (that are also taken with different VLA configurations)
shows an even higher percentage of steep spectrum sources.
This effect can be ascribed to a larger fraction of extended flux
detected with the more compact VLA configuration.
Possible biases introduced by the flat--radio-spectrum criterion in the 
radio-selected BL Lac samples cannot explain the discrepancies 
observed in the evolutionary properties of Radio and X-ray selected
samples of BL Lacs. 
}

\maketitle

\section{Introduction}
BL Lac objects are an enigmatic class of active galactic nuclei (AGN)
characterized by strong radio, optical and X-ray variability,
relatively high optical and radio polarization and featureless optical
spectra (e.g. Urry \& Padovani, 1995). The properties of the class can
be explained in terms of the {\it{relativistic beaming}} scenario
wherein the observed emission is dominated by Doppler-boosted
non-thermal radiation from a relativistic jet aligned with the line of
sight (Blandford \& Rees, 1978, Antonucci \& Ulvestad, 1985).

Owing to their featureless spectra and lack of UV excess in many BL Lacs, optical
search techniques (e.g. by excess colors, emission line strength),
used to search for other AGN, failed to find BL Lacs in large number
(e.g. Fleming et al., 1993). Since BL Lacs are both radio-loud and
X-ray-loud, surveys in these frequency bands discover them with high
efficiency (e.g. the ``1 Jansky'' sample: Stickel et al., 1991; the
{\it Einstein} Extended Medium-Sensitivity Survey (EMSS) sample:
Morris et al., 1991; Rector et al., 2000; the Deep X-ray Radio Blazar Survey (DXRBS) sample: Perlman et al., 1998, Landt et al. 2001; the ROSAT All Sky Survey (RASS) sample: Bade et al., 1998; the ``sedentary'' survey: Giommi,
Menna \& Padovani, 1999; 
the Radio Emitting X-ray (REX) survey sample: Caccianiga et al., 1999, Caccianiga et al., 2002). This splits BL Lacs into
two empirical subclasses, namely {\it radio-selected BL Lacs} (RBLs)
and {\it X-ray-selected BL Lacs} (XBLs).
Padovani \& Giommi (1995) proposed to distinguish these two classes on a more
physical basis, according to the spectral energy distribution (SED).
For ease of description here we will refer to RBL and XBL only.

RBLs and XBLs show different behaviors with respect to properties
like cosmological evolution, polarization, variability,
core-to-extended radio flux ratio, extended radio morphology,
spectral features (e.g. 
Wolter et al., 1994; Stickel et al., 1991
Morris et al., 1991;  Jannuzi, Smith
\& Elston, 1993, 1994; Perlman \& Stocke, 1993 to mention just a few) 
which have not been
satisfactorily explained yet. This led to the suggestion that the
observed discrepancies are at least partially caused by selection
effects due to different criteria used in radio and X-ray surveys.

In particular, the standard selection technique for RBLs requires a
flat-radio-spectrum criterion for which the radio spectral slope
must be less than 0.5 ($\alpha_r\leq0.5$; $S_\nu
\propto\nu^{-\alpha_r}$). This criterion has not been used for
classical X-ray selected samples of BL Lacs like the EMSS. 
If surveys with this
criterion select only part of the BL Lac population, the RBL samples
could be biased and the differences between RBLs and XBLs like the
ones mentioned above could be at least partially explained. 

It is already known that a number of objects with steep spectra
($\alpha_r>0.5$), classified as radio galaxies in radio surveys like
the ``1 Jansky'' (Owen, Ledlow, \& Keel, 1996, Perlman et al. 1996), 
otherwise meet the BL
Lacs selection criteria, based on optical properties (equivalent width
of emission lines; Ca~II break contrast) of March\~a et al. (1996), and
have broadband properties (radio luminosity and overall spectral energy 
distributions) that agree with those of BL Lacs (e.g. Rector et al., 2000).

To test if the flat--radio-spectrum criterion really introduces biases
in the BL Lacs selection and to estimate the incompleteness degree of RBLs
samples, we studied the radio spectral indices of a complete subsample
of XBLs in which the identification criteria were not based on the
radio spectral properties. 
Previous multi-frequency studies had already been performed on XBLs. For instance, Laurent-Muehleisen et al. (1993) collected non-\-simultaneous data for several objects of this class, that however do not make a complete sample.
We prefer to deal with the largest subsample for which we could have both
simultaneous and non-simultaneous data.

The sources of this subsample are extracted from the EMSS (Gioia et al., 1990;
Stocke et al., 1991), a 
catalog whose
BL Lacs sample has been thoroughly studied for completeness (Rector,
Stocke \& Perlman, 1999). A pilot study on the radio spectra of 8 XBLs 
from the EMSS and the {\it HEAO-1 A-2} all-sky survey (Piccinotti et al., 
1982) was
performed by Stocke et al. (1985). Three objects showed a spectral
slope which exceeds the limit for flat spectrum and two more
were marginally steep. Preliminary analysis of the EMSS BL Lacs
sample, by using non-simultaneous data at 6 and 20 cm from the EMSS
and the {\it NRAO VLA Sky Survey} (NVSS: Condon et al., 1998)
respectively, showed that about 30\% of the objects have a spectral
slope steeper than 0.5. This is the motivation that led us to investigate 
the EMSS sources by obtaining simultaneous spectra, which are not affected by 
variability.

We briefly describe the sample of objects used for this work in
\S\ref{The Sample}; in \S\ref{Data analysis} we present the data
reduction process as well as flux densities and radio spectra for each
object; in \S\ref{Results} we show the results obtained; our
conclusions are summarized in \S\ref{Conclusions}. Throughout the
paper we used $H_0=50$ km s$^{-1}$ Mpc$^{-1}$ and $q_0=0$, but
no conclusions are dependent upon this choice of cosmology.

\section{The Sample}
\label{The Sample}
\begin{table}[h]
\centering
{\footnotesize
\begin{tabular}{l c l}
\hline \hline
Name & Obs. date & z \\
\hline
MS0122.1+0903 & 16 Jun 2000 & 0.339$^*$ \\
MS0158.5+0019 & 16 Jun 2000 & 0.299$^*$ \\
MS0205.7+3509 & 16 Jun 2000 & $>0.351$$^+$ \\
MS0257.9+3429 & 16 Jun 2000 & 0.245$^*$ \\
MS0317.0+1834 & Jun 1984 & 0.190$^\dagger$ \\
MS0419.3+1943 & 16 Jun 2000 & 0.512$^*$ \\
MS0607.9+7108 & 16 Jun 2000 & 0.267$^*$ \\
MS0737.9+7441 & 16 Jun 2000 & 0.315$^*$ \\
MS0922.9+7459 & 13 Jun 2000 & 0.638$^*$ \\
MS0950.9+4929 & 13 Jun 2000 & ... \\
MS1019.0+5139 & 13 Jun 2000 & 0.141$^*$ \\
MS1050.7+4946 & 13 Jun 2000 & 0.140$^*$ \\
MS1207.9+3945 & Jun 1984 & 0.616$^\dagger$ \\
MS1221.8+2452 & 13 Jun 2000 & 0.218$^*$ \\
MS1229.2+6430 & 13 Jun 2000 & 0.164$^*$ \\
MS1235.4+6315 & Jun 1984 & 0.297$^\dagger$ \\
MS1402.3+0416 & Jun 1984 & 0.344:$^\dagger$ \\
MS1407.9+5954 & 13 Jun 2000 & 0.495$^*$ \\
MS1443.5+6348 & 13 Jun 2000 & 0.299$^*$ \\
MS1458.8+2249 & 13 Jun 2000 & 0.235$^*$ \\
MS1534.2+0148 & 13 Jun 2000 & 0.312$^*$ \\
MS1552.1+2020 & 13 Jun 2000 & 0.222$^*$ \\
MS1757.7+7034 & 13 Jun 2000 & 0.407$^*$ \\
MS2143.4+0704 & 16 Jun 2000 & 0.237$^*$ \\
MS2336.5+0517 & 16 Jun 2000 & 0.74::$^\dagger$ \\
MS2347.4+1924 & 16 Jun 2000 & 0.515$^*$ \\
\hline \hline
\multicolumn{3}{l}{($^*$)  \scriptsize\it Redshift from Morris et al. (1991)} \\
\multicolumn{3}{l}{($^+$)  \scriptsize\it Redshift from Watson et al. (2004)} \\
\multicolumn{3}{l}{($^\dagger$)  \scriptsize\it Redshift from Rector et al. (2000)} \\
\end{tabular}
}
\caption{The CWSR sample.}
\label{list sources}
\end{table}

We restrict the analysis to the EMSS north of $\delta=-20^\circ$ in
order to ensure 100\% identification rate. The revised EMSS BL Lac
sample (Rector et al., 2000) then contains 36 BL Lacs from which we
observed a subsample of 22 objects, named hereafter CWSR. The sources
were selected imposing f$_{6 cm} \geq 1$ mJy in the EMSS dataset and
celestial positions that allowed observations at the Very Large Array 
(VLA\footnote{The VLA is a facility of the National Radio Astronomy Observatory. The National Radio Astronomy Observatory (NRAO) is a facility of the National Science Foundation operated under cooperative agreement by Associated Universities,
Inc.}) 
during June 2000. The resulting subsample is therefore not biased with 
respect to radio spectral index.

The VLA observations where performed on June 13 and June 16 in the L 
($\lambda=20$ cm), C ($\lambda=6$ cm) and X ($\lambda=3.6$ cm) frequency 
bands. 
The observations were carried out in the C configuration.
Thus, the maps have an angular resolution equal
to 16$^{\prime\prime}$.3, 5$^{\prime\prime}$.1 and 3$^{\prime\prime}$ 
at 20, 6 and 3.6 cm respectively.
Each object was observed for $\sim$10 minutes per frequency, obtaining
a sensitivity of about 0.05 mJy beam$^{-1}$.
Flux densities were bootstrapped from 3C 286 (June 13) and 3C 48 (June
16) (Baars et al. 1977) 

If we include in the CWSR sample the 4 EMSS objects published in Stocke et al. (1985),
observed simultaneously on June 1984 at all the available wavelengths (20, 18, 6, 2.0 and 1.3 cm) with the VLA in the C/D hybrid configuration, we have 26
sources in total.
These sources are listed in Table \ref{list sources}:
the first column indicates the source names, the second one
the observation dates and the third one the redshift values.

For the same 26 objects 
we also have non-\-simultaneous data at 20 and 6 cm, respectively from
NVSS and EMSS archives.

\section{Data analysis}
\label{Data analysis}

\subsection{Radio maps}
\label{Radio map}

\begin{table*}
\centering
{\scriptsize
\begin{tabular}{|c||r|r|r||r|r|r||r|r|r|}
\hline
& \multicolumn{3}{c||}{$L$ band} & \multicolumn{3}{c||}{$C$ band} & \multicolumn{3}{c|}{$X$ band}\\
\cline{2-10}
Name & $\;F_{tot}${\scriptsize $^a\;$} & $\;F_{peak}${\scriptsize $^a\;$} & rms($\times 10^{-2}$){\scriptsize $^b$} & $\;F_{tot}${\scriptsize $^a\;$} & $\;F_{peak}${\scriptsize $^a\;$} & rms($\times 10^{-2}$){\scriptsize $^b$} & $\;F_{tot}${\scriptsize $^

a\;$} & $\;F_{peak}${\scriptsize $^a\;$} & rms($\times 10^{-2}$){\scriptsize $^b$} \\
\hline \hline
MS0122.1+0903 & 1.34 & 1.44 & 6.26 & 1.47 & 1.39 & 3.30 & 1.28 & 1.08 & 2.60 \\
MS0158.5+0019 & 13.26 & 12.46 & 8.67 & 9.70 & 8.87 & 3.00 & 8.16 & 7.36 & 3.00 \\
MS0205.7+3509 & 3.53 & 3.53 & 13.85 & 5.24 & 5.04 & 3.40 & 4.96 & 4.24 & 2.49 \\
MS0257.9+3429 & 8.80 & 8.23 & 10.00 & 10.79 & 9.41 & 3.80 & 11.23 & 9.98 & 3.05 \\
MS0419.3+1943 & 6.10 & 5.66 & 8.44 & 9.05 & 7.91 & 3.18 & 8.07 & 7.16 & 2.57 \\ 
MS0607.9+7108 & 15.33 & 13.12 & 40.71 & 14.37 & 13.97 & 3.58 & 10.34 & 9.82 & 3.74 \\
MS0737.9+7441 & 25.60 & 24.78 & 9.29 & 22.46 & 21.04 & 3.45 & 20.07 & 19.39 & 3.26 \\
MS0922.9+7459 & 43.37 & 21.88 & 8.57 & 4.17 & 3.59 & 4.00 & 2.89 & 2.66 & 3.30 \\
MS0950.9+4929 & 4.58 & 4.56 & 5.67 & 3.42 & 3.16 & 2.88 & 3.25 & 2.89 & 2.72 \\
MS1019.0+5139 & 2.71 & 2.84 & 8.03 & 2.97 & 2.75 & 3.16 & 2.86 & 2.70 & 2.60 \\
MS1050.7+4946 & 57.13 & 48.84 & 7.34 & 42.71 & 39.59 & 3.46 & 38.19 & 32.44 & 3.31 \\
MS1221.8+2452 & 23.31 & 21.17 & 13.02 & 22.00 & 20.95 & 3.50 & 21.59 & 19.80 & 2.75 \\
MS1229.2+6430 & 54.68 & 51.51 & 11.87 & 46.31 & 43.97 & 4.65 & 40.96 & 38.21 & 3.20 \\
MS1407.9+5954 & 31.51 & 24.54 & 12.67 & 19.55 & 16.80 & 4.44 & 16.77 & 13.67 & 3.02 \\
MS1443.5+6348 & 9.74 & 9.55 & 51.29 & 4.47 & 3.88 & 4.69 & 6.29 & 5.11 & 3.17 \\
MS1458.8+2249 & 48.42 & 47.47 & 8.75 & 87.99 & 85.33 & 5.23 & 94.41 & 91.87 & 3.00 \\
MS1534.2+0148 & 65.12 & 60.16 & 8.10 & 34.07 & 25.29 & 3.29 & 24.01 & 15.96 & 2.57 \\
MS1552.1+2020 & 64.46 & 56.92 & 53.89 & 38.79 & 31.93 & 3.79 & 29.61 & 24.06 & 2.71 \\ 
MS1757.7+7034 & 11.68 & 11.45 & 15.61 & 11.15 & 10.53 & 3.78 & 9.80 & 9.03 & 3.07 \\
MS2143.4+0704 & 78.59 & 65.10 & 15.48 & 51.87 & 45.27 & 6.31 & 43.57 & 37.21 & 2.29 \\
MS2336.5+0517 & 11.82 & 8.83 & 12.62 & 5.86 & 5.26 & 3.30 & 4.69 & 4.38 & 2.80 \\
MS2347.4+1924 & 4.27 & 4.64 & 9.96 & 3.01 & 2.87 & 2.90 & 2.42 & 2.19 & 2.70 \\
\hline
\multicolumn{10}{l}{$^a$  \scriptsize\it Total and Peak Flux Density in mJy} \\
\multicolumn{10}{l}{$^b$  \scriptsize\it Root Mean Square in mJy $\times$ beam$^{-1}$} \\ 
\end{tabular}
}
\caption{$F_{tot}$, $F_{peak}$ and rms measured in the L (20 cm), C (6 cm) e X (3.6 cm) frequency band for the 22 XBLs of CWSR sample observed on June 2000.}
\label{Sim.fluxes}
\end{table*}

Radio maps for the 22 CWSR sources observed on June 2000
were produced in $\mathcal{AIPS}$ with a standard procedure
that includes: removal of RFI from (u,v) data caused by incidental
factors during the observations, calibration of all sources in
amplitude and in phase, creation of a ``cleaned'' map of brightness
distribution by using a CLEAN algorithm. Particular care was taken at
20cm to look and correct for contaminating sources.

For a more stable phase calibration, a calibrator source was observed 
before and after each source observation.
Whenever possible, self-calibration was also applied.

We adopted a uniform weight function. We did some comparison with maps 
derived from natural weighting function and no significant differences 
were observed.

The resulting maps do not show morphological structures that are worth
publishing. All sources are fitted by Gaussians (even if sometimes with 
widths larger than the beam size).

\subsection{Flux densities}
\subsubsection{Simultaneous flux densities}
\label{SFD}

\begin{table*}
\centering
{\scriptsize
\begin{tabular}{|c||r|r|r|r|r|}
\hline
& \multicolumn{5}{c|}{$F_{tot}$ (mJy)}\\
\cline{2-6}
Name & 20cm & 18cm & 6cm & 2cm & 1.3cm \\
\hline \hline
MS0317.0+1834 & 17.3$\pm$0.6 & 16.5$\pm$0.8 & 11.5$\pm$0.23 & 5.5$\pm$0.3 & $<$4.5 \\
MS1207.9+3945 & 15.1$\pm$1.1 & 18.3$\pm$1.3 & 6.1$\pm$0.4 & 0.9$\pm$0.3 & $<$4.2 \\
MS1235.4+6315 & 10.7$\pm$0.5 & 10.1$\pm$0.5 & 7.0$\pm$0.38 & 4.6$\pm$0.3 & {\scriptsize Not Obs.} \\ 
MS1402.3+0416 & 35.5$\pm$1.3 & 38.2$\pm$1.4 & 20.8$\pm$0.4 & 14.5$\pm$0.3 & 8.7$\pm$1.4 \\
\hline
\end{tabular}
}
\caption{$F_{tot}$ and rms reported from Stocke et al. (1985) for the 4 EMSS objects included in the CWSR sample.}
\label{Stocke}
\end{table*}

From the ``cleaned'' maps we computed flux densities and their
associated rms errors by assuming the source signal as a Gaussian.  
We measured both the peak flux density
($F_{peak}$) and the total flux density ($F_{tot}$).
We assume that $F_{peak}$ represents the emission from the source core
and we define the extended flux $F_{ext}$ as the difference between 
$F_{tot}$ and $F_{peak}$.

In Table \ref{Sim.fluxes} $F_{tot}$, $F_{peak}$ and their associated
rms errors are listed for the CWSR sample.  By comparing the peak and
total flux densities, we confirm  that this sample and
the C data we obtained follow the BL Lac trend to be compact: we
found that the maximum value of the ratio $F_{tot}/F_{peak}$ is 1.98 at 20
cm. In \S\ref{ctratio} we better analyze this aspect.

The noise for each map was determined from the rms fluctuations in
areas free of radio emission. We compared these rms errors (see Table
\ref{Sim.fluxes}) with their ideal values (0.032, 0.028 and 0.023 mJy in 
L, C and X band respectively) from
the VLA manual (Perley, 2000).
While in C and X bands the rms errors were found similar to their
ideal limits, in L band they were larger than their theoretical
values. This can be explained by additional interference at 20 cm, and
is not a particularly unexpected result.

For 3 sources (MS0122.1+0903, MS1019.0+5139, MS2\-347+1924) at 20 cm
we measured a total flux density smaller than the peak flux density
(but consistent within the errors)
probably because of phase errors not completely removed from radio
maps by cleaning processes. Due to the small numerical difference
between $F_{tot}$ and $F_{peak}$ ($F_{tot}/F_{peak}$ equal to 0.93,
0.95 and 0.92 respectively) and since BL Lacs are known as objects with
a high core-to-extended ratio (e.g. Perlman \& Stocke, 1993), this
does not introduce significant biases.

In three sources (namely MS0607.9+7108, MS1443.5+6348, MS1552+2020) we 
measured a very high rms threshold 
because of a source in the neighborhood, either very bright or 
extended, that could not be properly cleaned. 
In Table \ref{Stocke} we report simultaneous data for the 4  objects included
in the CWSR sample from Stocke et al. (1985). The
total flux density $F_{tot}$ and their associated rms errors are
listed at all the observed wavelengths.

\subsubsection{Non-\-simultaneous flux densities}

\begin{table}[!h]
\centering
{\scriptsize
\begin{tabular}{|l||r|r||r|}
\hline
& \multicolumn{2}{c||}{NVSS $^*$}& EMSS $^*$  \\
& \multicolumn{2}{c||}{(20cm)} & (6cm) \\
\cline{2-4}
Name & $F_{tot}$ &$F_{peak}$ & $F_{tot}$ \\ 
\hline \hline
MS0122.1+0903 & 2.49 &1.75 & 1.4 \\
MS0158.5+0019 & 12.7 &12.7 & 11.3 \\
MS0205.7+3509 & 4.7 &4.7 & 3.6 \\
MS0257.9+3429 & 10.1 &10.1 & 10.0 \\	
MS0317.0+1834 & 20.9 &17.9 & 11.5 \\   
MS0419.3+1943 & 2.85 &2.76 & 8.0 \\	
MS0607.9+7108 & 25.2 &25.2 & 18.2 \\		
MS0737.9+7441 & 22.7 &22.7 & 24.0 \\		
MS0922.9+7459 & 84.2 &59.2 & 3.3 \\		
MS0950.9+4929 & 2.6 &2.6 & 3.3 \\
MS1019.0+5139 & 4.8 &4.8 & 2.4 \\	
MS1050.7+4946 & 64.1 &64.1 & 53.8 \\		
MS1207.9+3945 & 18.7 &18.7 & 6.1 \\		
MS1221.8+2452 & 24.5 &24.5 & 26.4 \\	
MS1229.2+6430 & 58.0 &54.5 & 42.0 \\	
MS1235.4+6315 & 12.1 &12.1 & 7.0 \\		
MS1402.3+0416 & 40.7 &40.7 & 20.8 \\		
MS1407.9+5954 & 35.7 &33.3 & 16.5 \\		
MS1443.5+6348 & 17.6 &17.6 & 11.6 \\	
MS1458.8+2249 & 31.9 &31.9 & 29.8 \\		
MS1534.2+0148 & 72.3 &72.3 & 34.0 \\		
MS1552.1+2020 & 79.0 &79.0 & 37.5 \\		
MS1757.7+7034 & 10.3 &10.3 & 7.2 \\		
MS2143.4+0704 & 101.1 &95.1 & 50.0 \\
MS2336.5+0517 & 16.5 &13.6 & 4.9 \\		
MS2347.4+1924 & 4.7 &4.7 & 3.2 \\
\hline
\multicolumn{3}{l}{$^*$ \scriptsize\it Total and Peak Flux Density in mJy} \\
\end{tabular}
}
\caption{Total flux densities drawn from NVSS (20 cm) and EMSS (6 cm) archives for the 26 
sources. NVSS peak flux density is also listed.}
\label{Non-sim.fluxes}
\end{table}

Most datasets available to compute radio spectra are not
simultaneous; moreover, BL Lacs are known to be variable objects
(e.g. Angel \& Stockman, 1980; Ulrich, Maraschi \& Urry, 1995). To
assess the impact of variability on non-\-simultaneous measurements of
$\alpha_r$, we statistically compare the $\alpha_r$ values measured from
simultaneous data to values determined with non-\-simultaneous
data. We used 20cm NVSS (Condon et al. 1998) and 6cm EMSS (Maccacaro et
al. 1994) catalog fluxes.
Since MS0122.1+0903 and MS0419.3+1943 do not appear in the NVSS catalog
(they are at the completeness limit)
we measured their fluxes from the  NVSS on-line radio
images which can be down-loaded in FITS format (see {\it
http://www.cv.nrao.edu/NVSS/postage.html}).

In Table \ref{Non-sim.fluxes} flux densities are shown for the 26 
listed objects.

\subsection{Radio Spectra}
\label{Radio Spectra}

\begin{table}[!ht]
\centering
{\scriptsize
\begin{tabular}{|l||r||r|r|r|}
\hline
 & sim. $\alpha_r$ & \multicolumn{3}{c|}{non-sim. $\alpha_r$} \\
\cline{2-5}
Name & \tiny{CWSR} & \tiny{NVSS/} & \tiny{NVSS/} & \tiny{CWSR/} \\
Name & -- & \tiny{EMSS} & \tiny{CWSR} & \tiny{EMSS} \\
\hline \hline
MS0122.1+0903 & 0.09 & 0.48 & 0.44 & 0.02 \\
MS0158.5+0019 & 0.28 & 0.10 & 0.22 & 0.13 \\
MS0205.7+3509 & $-$0.22 & 0.22 & $-$0.09 & $-$0.017 \\
MS0257.9+3429 & $-$0.15 & 0.01 & 0.05 & $-$0.11 \\	
MS0317.0+1834 & 0.48 & 0.57 & 0.50 & 0.02 \\   
MS0419.3+1943 & $-$0.19 & $-$0.86 & $-$1.00 & $-$0.23 \\	
MS0607.9+7108 & 0.20 & 0.27 & 0.47 & $-$0.14 \\		
MS0737.9+7441 & 0.14 & $-$0.05 & 0.01 & 0.05 \\		
MS0922.9+7459 & 0.72 & 2.70 & 2.50 & 2.32 \\		
MS0950.9+4929 & 0.21 & $-$0.20 & $-$0.23 & 0.27 \\
MS1019.0+5139 & $-$0.01 & 0.56 & 0.40 & 0.14 \\	
MS1050.7+4946 & 0.24 & 0.15 & 0.34 & 0.05 \\		
MS1207.9+3945 & 1.25 & 0.94 & 0.93 & 0.80 \\		
MS1221.8+2452 & 0.05 & $-$0.06 & 0.09 & $-$0.10 \\	
MS1229.2+6430 & 0.16 & 0.27 & 0.19 & 0.22 \\	
MS1235.4+6315 & 0.36 & 0.48 & 0.46 & 0.35 \\		
MS1402.3+0416 & 0.42 & 0.59 & 0.56 & 0.64 \\		
MS1407.9+5954 & 0.37 & 0.64 & 0.50 & 0.54 \\		
MS1443.5+6348 & 0.33 & 0.35 & 1.14 & $-$0.15 \\	
MS1458.8+2249 & $-$0.41 & 0.06 & $-$0.84 & 0.40 \\		
MS1534.2+0148 & 0.57 & 0.63 & 0.63 & 0.54 \\		
MS1552.1+2020 & 0.45 & 0.62 & 0.59 & 0.45 \\		
MS1757.7+7034 & 0.09 & 0.30 & $-$0.07 & 0.40 \\		
MS2143.4+0704 & 0.34 & 0.59 & 0.55 & 0.38 \\
MS2336.5+0517 & 0.53 & 1.01 & 0.86 & 0.73 \\		
MS2347.4+1924 & 0.39 & 0.32 & 0.37 & 0.31 \\
\hline
\end{tabular}
}
\caption{Simultaneous and non-\-simultaneous spectral indices are listed for 26 objects ($S_{\nu}\propto\nu^{-\alpha_r}$).}
\label{slopes}
\end{table}

By using the multi-frequency flux densities measured (Tables
\ref{Sim.fluxes} and \ref{Stocke}) we construct simultaneous spectra for CWSR sources. 
The spectral shape is assumed to be a power law ($S_\nu\propto\nu^{-\alpha_r}$). 
The slopes were computed with the Least Square
Method (LSQ) and are reported in Table \ref{slopes}.
At 20 cm the MS0922.9+7459 radio map is confused due to the
superposition of another source (probably a radio galaxy in a
foreground cluster). Thus, the corresponding flux density reported in
this work, even if it is the best measure we could obtain, is to be
considered as an upper limit and was not included in all the spectral
slope calculations for this source. 

To investigate whether variability could affect the spectral slope, we
calculated simultaneous and non-simultaneous two-point (20 cm + 6
cm) $\alpha_r$ values with the EMSS (6 cm), NVSS (20 cm) and June 2000
(20 cm + 6 cm) observations for the CWSR sources.

We computed as well the spectral index for the source cores,
finding no significant changes with respect the $\alpha_r$
calculated with total flux density. 
The only source that presents a remarkable difference between the core 
spectral index ($\alpha_r=0.14$) and the total spectral index 
($\alpha_r \leq 0.72$) is MS0922.9+7459.

In Table \ref{slopes}, spectral indices are listed from
non-\-simultaneous and simultaneous spectra. The second column
gives simultaneous slopes for the 26 sources of the CWSR sample;
in the third, fourth and fifth column
non-\-simultaneous slopes are computed,
respectively, with 20cm NVSS and 6cm EMSS observations, with 20cm NVSS
and 6cm CWSR observations, with 20cm CWSR and 6cm EMSS observations. 
By comparing different values of spectral slope, we note that 11 of 26 objects
have spectral indices that could be classified differently
depending on the dataset used; in
particularly, 9 of 11 sources could have a flat or steep spectrum, 1 source 
(MS0419+19\-43) could have an inverted or flat spectrum, 1 source (MS1458.8+2249) 
could have an inverted, flat or steep spectrum.

In the next section we will consider a few explanation that could affect
these measurements:
the well-known variability of these objects and
the angular resolution of the observations. Data taken with different angular
resolution in fact could measure, if present, a different amount of extended
flux.

\section{Results} 
\label{Results}

\subsection{Core-to-total flux density ratio}
\label{ctratio}

\begin{table}[hb]
\centering
{\footnotesize
\begin{tabular}{|l|r|r|}
\hline
frequency band & $\alpha$ & $\beta$ \\
\hline \hline
$L$ (20 cm) & 0.94 & 0.02 \\
$L$ (20 cm) $^*$ & 0.96 & 0.01 \\
$C$ (6 cm) & 0.99 & $-$0.03 \\
$X$ (3.6 cm) & 1.00 & $-$0.06 \\
\hline
\multicolumn{3}{l}{$^*$\scriptsize\it Fit without MS0922.9+7459.} \\
\end{tabular}
}
\caption{Results of the fit of F$_{Peak}$ vs. F$_{Tot}$ for the CWSR sources
indicating that an extended component is visible at 20 cm.}
\label{lsq feat.}
\end{table}

It is already known that BL Lacs are very core-dominated sources
(e.g. Perlman \& Stocke, 1993) and that this behavior is more
evident in higher frequency observations. This is confirmed from our data 
(see section \ref{SFD}).
In fact, we confirm the trend of the core-dominance in the CWSR sample by
fitting $F_{peak}$ vs $F_{tot}$ (from Table~\ref{Sim.fluxes}) with the LSQ. 
The LSQ slope and intercept are given in Table \ref{lsq
feat.}: the first column gives the VLA frequency band, the second and
the third column, respectively, give the slope and the intercept of
the LSQ fit. 
We compute the fit also excluding the L band value of MS0922.9+7459 (that is confused, see section \ref{Radio Spectra})

The errors on the slopes are typically 0.1-0.2, and therefore only 
at 20 cm there is a small deviation
from a linear fit, indicating the presence of an extended component which is
more visible for brighter sources.

We also checked for a possible redshift dependence of the core-dominance
since extended surface brightness dims
as $(1+z)^{-4}$, so significant amount of flux from the
extended component could be lost at high z. In addition, since at large
distances a given angular resolution corresponds to a larger physical
size, if much of the extended flux of a source is close to the core,
some of its inner part could be subsumed into the central component
and therefore included with the measured core flux. Perlman \& Stocke
(1993) studied the possible biases introduced by
these terms and found that for sources with $z\leq0.5$ there is only a
modest correction. Since the objects in this work are all at
$z\mincir0.5$, we estimate that the loss of the extended flux
due to these biases was not significant and, thus, was not considered.

\subsection{Spectral indices}

\begin{figure}[!h]
\centering
\includegraphics[width=8cm]{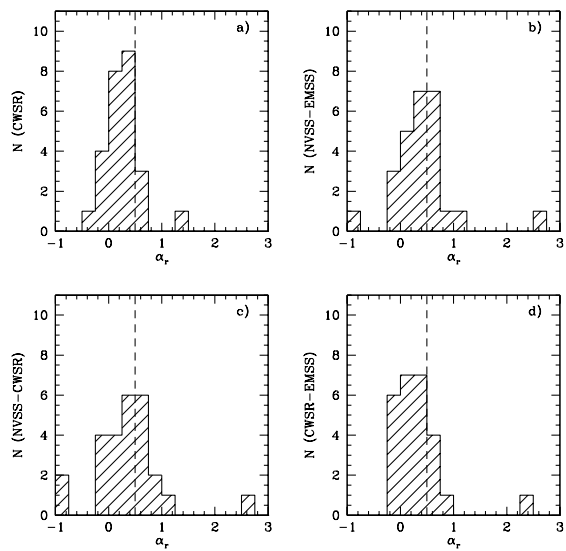}
\caption{The $\alpha_r$ distributions for the 26 sources in the CWSR sample.
Plot a), upper left, shows the spectral indices calculated from the 
simultaneous CWSR data. Plot b), upper right, shows the spectral indices calculated from 20cm NVSS data and 6cm EMSS data. Plot c), bottom left, shows the spectral indices calculated from 20cm NVSS data and 6cm CWSR data. Plot d), bottom right, shows the spectral indices calculated from 20cm CWSR data and 6cm EMSS data. The vertical dashed line represents the cut-off value $\alpha_r=0.5$.}
\label{histall}
\end{figure}

\begin{table*}[!ht]
\centering
{\scriptsize
\begin{tabular}{|l r|c|c|c|}
\hline
\multicolumn{2}{|c|}{datasets} & \% & $\Delta$ & VLA \\
20 cm & 6 cm & steep sp. & yrs & config \\
\hline \hline 
CWSR & CWSR & 15.4 & 0 &C+C/D$^*$\\
\hline
NVSS & EMSS & 38.5 & 5-10 &D, C\\
NVSS & CWSR & 38.5 & 4-7 &D, C\\
CWSR & EMSS & 23.1 & 12-14 &C+C/D, C\\
\hline 
\multicolumn{5}{l}{$^*$ \scriptsize\it Also 3.6 cm observations in C configuration and 18 and 2 cm observations} \\
\multicolumn{5}{l}{\,\,\,\, \scriptsize\it in C/D configuration are used} \\
\end{tabular}
}
\caption{The percentage of objects with steep spectra ($\alpha_r>0.5$; $S_{\nu}\propto\nu^{-\alpha_r}$) is listed for different datasets.}
\label{perc. steep spectra}
\end{table*}

We studied the distribution of spectral slopes reported in Table~\ref{slopes}.
In Figure \ref{histall} 
the 4 $\alpha_r$ distributions obtained for the simultaneous and
non-\-simultaneous datasets are shown. The dividing line at $\alpha_r$ = 0.5
is plotted as a dashed line.

In Table \ref{perc. steep spectra}  we report the percentage of sources
with a spectrum steeper than $\alpha_r$ = 0.5.
The first column gives the
different datasets used to compute the spectra; 
the second column gives the
percentage of steep spectra computed; the third column gives the time
span between the different datasets; the fourth column gives the VLA
configurations used for the observations.
As shown in Table \ref{perc. steep spectra}, we found that about 15\%
of the sources have steep spectra computed with simultaneous data, and even
larger percentages with non-simultaneous data.
As shown in Figure~\ref{histall} the distribution of  $\alpha_r$ is a
rather continuous one, so that picking a single value to divide it in two
populations is rather arbitrary, and does not have a physical meaning.
However, this has both practical and historical reasons (see e.g. the 1Jy 
BL Lac sample definition in K\"uhr \& Schmidt, 1990) and so it remains
an important issue to be addressed.

To investigate whether the different subsamples under consideration
are consistent with each other, we performed a Kolmogorov-Smirnov test
for the four $\alpha_r$ distributions (see Figure \ref{histall}) by
comparing their respective datasets considered in Table
\ref{perc. steep spectra}. We cannot exclude that the four dataset 
combinations are extracted from the same population (P$>$32\%).
However, the small number of points does not allow any further investigations.

Since by using non-simultaneous data we find a higher
percentage of steep spectrum sources, it seems straightforward to ascribe
this effect to flux variability, that we know is present in this class
of objects. 
However, other factors are at play.
Again from Table \ref{perc. steep spectra}, we notice that the higher 
percentages of steep spectra correspond to the use of the 20cm NVSS data, 
which were carried out with the VLA array in D configuration. 
As previously discussed (see
\ref{ctratio}), 
this probably indicates that a larger
fraction of extended flux, which has a spectrum steeper than the core
emission, is detected in these datasets.
The Gaussian distribution of the NVSS-to-CWSR 20cm
data ratio seems to confirm this
hypothesis: the distribution mean value is $x_{mean}=1.19$, indicating
that the NVSS flux densities exceed on average the CWSR values by 19\%.

The effect cannot be ascribed to confusion due to a large beam,
since inspection of the higher resolution C configuration images shows
that no sources fall into the NVSS beam beside the target BL Lac.

Thus, along with variability, we might suggest that a significant 
fraction of extended flux, that is considered negligible by BL Lacs 
beaming models (e.g. Blandford \& Rees, 1978, Antonucci \& Ulvestad, 1985), is present, 
and could at least partially explain the differences obtained in the
$\alpha_r$ distributions.

\begin{figure}[!ht]
\centering
\includegraphics[width=6cm]{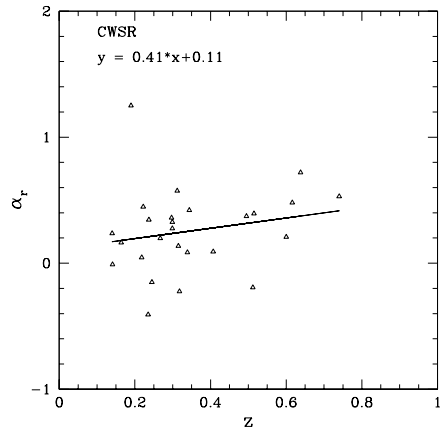}
\caption{$\alpha_r$ against z is plotted for the simultaneous dataset. The LSQ fit and the respective functional form are also shown.}
\label{zalfa}
\end{figure}

To investigate what influence the detection of an extended component has
on the spectral index, 
we plotted the spectral slopes of the CWSR sources against redshift z
in Figure \ref{zalfa}.
We confirm what we assumed in \S 4.1 that no redshift effect is
present in the detection of extended flux due to decreasing of brightness.
The LSQ fit in Figure \ref{zalfa} is consistent with a constant
due to the large error on slope (0.41$\pm$ 0.39)
and if anything sources at larger redshift seem to have steeper slopes.

To assess possible biases due to the varying spectral slope, we
computed total luminosity for the 25 XBLs with known redshift in the
L, C and X frequency bands.  For the 22 CWSR sources observed on June
2000, peak and extended luminosities could also be computed from the
peak and total flux densities; where extended emission was not
detected, conservative upper limits on extended radio luminosity
levels were obtained by assuming that each source has uniformly bright
extended emission at 1$\sigma$ detection level over a 3000 kpc$^2$
area surrounding the core (see Rector et al. 2000 for a description of
how to compute the extended luminosities).  The luminosity of the
extended component is in general in agreement with what found
previously (e.g. Rector et al. 2000, Perlman \& Stocke 1993) so we do
not report values here. In fact the C configuration snapshot data do
not have a sufficient {\it uv} coverage to warrant a better definition
of the core than the previous A configuration data on the same
sources.

To test whether there is a correlation between luminosities and
spectral slope, we computed the correlation coefficient for the LSQ
fit and its goodness probability. 
We found again that there is a 99\% probability that a correlation exists 
only for extended luminosity at 20cm.
We check also a possible dependence of $\alpha_r$ on NVSS and EMSS 
radio luminosities.
The correlation between spectral slope and 20cm luminosity seems to
support the idea that the extended component is important in
these objects and that this is the main factor to explain the
difference in the spectral measures.

\begin{figure}[!ht]
\centering
\includegraphics[width=7.5cm]{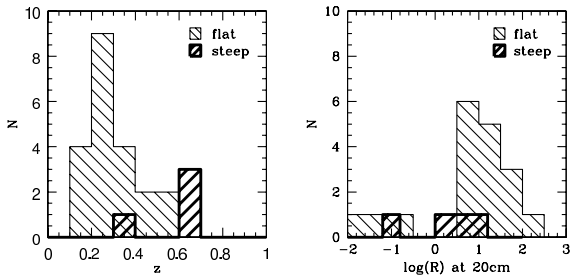}
\caption{The redshift {\it left} and core dominance {\it right} distribution of flat spectra objects compared to the steep spectra ones.}
\label{zrfns}
\end{figure}

By splitting our sample of 26 objects in two groups, 
steep and flat spectra, we studied the different
distribution of redshift, radio luminosity (at 20 and 6 cm), X
luminosity (0.3-3.5 keV) and 20cm core dominance.

We plot in Figure \ref{zrfns}{\it (left)} the histogram of the redshift distribution
for objects with $\alpha_r >$ 0.5 and $\alpha_r \leq $ 0.5 respectively.
The flat-spectrum dataset has on average redshift
values smaller than the steep-spectrum one. Since this is a flux limited
sample, it is fair that we find the same behavior 
also for the radio and X-ray luminosities.
However we stress that the redshift range of
this sample is small, and in particular lower than the radio selected
samples like the 1Jy.

In Figure \ref{zrfns}{\it (right)}, we show the histogram of core dominance 
for the two datasets, at 20 cm.
As expected, the steep spectrum sources show in general a smaller degree
of core dominance, even if the statistical significance of the result
is marginal.

\begin{figure}[!ht]
\centering
\includegraphics[width=6cm]{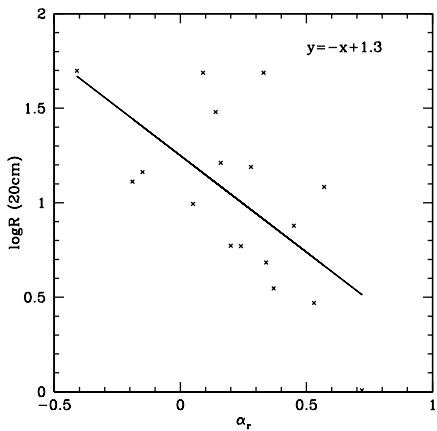}
\caption{The logarithm of core-dominance R (at 20 cm) is plotted against
the radio spectrum $\alpha_r$. The regression line has a slope of $-1.02$
with rather large uncertainties (0.32).}
\label{acd}
\end{figure}

We also plot in Figure~\ref{acd} the core-dominance at 20 cm against the
radio spectral slope  $\alpha_r$, for the sources for which we have a measure
(i.e. not an upper limit) for R. The two variables have a correlation 
coefficient of $-0.63$, indicating that, within the available statistics,
they are correlated at $>$ 99\%.

\subsection{$\langle V_e/V_a \rangle$ and the spectral index selection}

\begin{figure}[!ht]
\centering
\includegraphics[width=6cm]{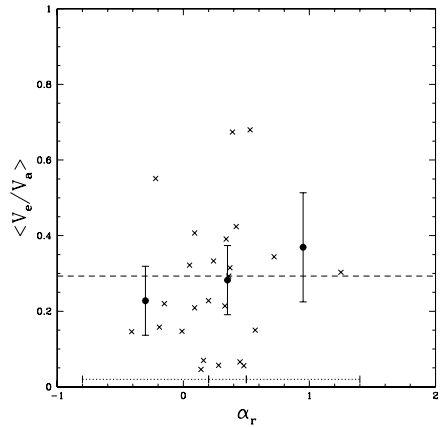}
\caption{$V_e/V_a$ against $\alpha_r$  is shown for the 25 XBLs with
a redshift measure. 
The $\langle V_e/V_a \rangle$ trend is also plotted with error-bars for 
$\alpha_r$
grouped in three bins ($\alpha_r<0.2$; $0.2<\alpha_r<0.5$;
$\alpha_r>0.5$). The dotted line represents the mean value of 
$\langle V_e/V_a \rangle$.}
\label{arvmf}
\end{figure}

One controversial issue in the study of BL Lacs is their evolutionary
behavior.
The first studies were based on the EMSS and found evidence for ``negative
evolution'' (i.e. less numerous/luminous objects in the past; Maccacaro et al.
1984; Morris et al. 1991; Wolter et al. 1994; Rector et al. 2000).
At the same time the 1Jy sample (selected in the radio band, and with
a constraint on  $\alpha_r<0.5$, K\"uhr \& Schmidt 1990) showed no or mild
positive evolution (Stickel et al. 1991; Rector \& Stocke 2001).
It is therefore interesting to think that the different criteria,
in particular the flat radio spectral index, could select samples with
different evolution properties. This result is
obtained if the steep spectrum BL Lacs, present only in X-ray selected
samples, all show a very negative evolution.
Since the number involved in these samples are small (20-40 objects
in total), even few objects have a large weight.

We used the standard $\langle V_e/V_a \rangle$ test (Avni \& Bachall, 1980)
to check the uniformity of the spatial distribution of the objects
in a flux limited sample. For each object the two volumes ($V_e$
and $V_a$) are derived
on the basis of the observed redshift and the maximum redshift at which
the object could be still included in the survey. A value of 
$\langle V_e/V_a \rangle$ = 0.5 is given by a uniform distribution,
while lower values of $\langle V_e/V_a \rangle$ indicate a distribution 
skewed towards low redshifts. Details can be found
e.g. in Morris et al (1991). We stress here that only the X-ray flux limits 
are used to derive the maximum available volume.
In fact the EMSS did not have any requirement on the
radio flux or optical flux to have an object enter the sample, and all
the EMSS BL Lacs have been also detected in the radio band.

To test the impact of the $\alpha_r$ selection criteria on the measure
of the evolution, we studied the correlation between
$V_e/V_a$ and the radio spectral index $\alpha_r$ for the 25 objects 
in the CWSR sample with a redshift measure (see Rector et al. 2000 for the 
$V_e/V_a$ values), plotted in Figure \ref{arvmf}.
The data have been also grouped in three  $\alpha_r$ bins to better show the
trend 
($\alpha_r<0.2$; $0.2<\alpha_r<0.5$;
$\alpha_r>0.5$) and plotted in  Figure \ref{arvmf} with error bars
that reflect number statistics. 
We find that the mean value is consistent with a constant
in the 3 intervals of $\alpha_r$. If anything there is a slight suggestion
that XBLs with steep spectra ($\alpha_r>0.5$) have a higher mean
$\langle V_e/V_a \rangle$ value.
Thus, it seems that the exclusion of steep spectra objects in RBLs samples
could not cause the differences observed in the cosmological evolution
with respect to the EMSS sample. 

The picture is however far from clear: while surveys like the Deep
X-ray Radio Blazar Survey (DXRBS; Perlman et al. 1998; Padovani et
al. 2002; Giommi et al., 2002), the ``sedentary'' survey (Giommi et
al. 1999) and the ROSAT - All Sky Survey (RASS; Bade et al. 1998;
Beckmann et al.  2003) found very low values of $\langle V_e/V_a
\rangle$ for XBLs, other surveys like the Radio Emitting X-ray survey
(REX; Caccianiga et al. 1999, Caccianiga et al., 2002) did not find
the same result, possibly indicating that the negative evolution in
this class of objects is not so strong as was previously measured.

\section{Conclusions}
\label{Conclusions}
By using fluxes at 20, 6 and 3.6 cm, simultaneously measured with the
VLA, we computed the radio spectra for a complete subsample of 22 XBLs
from the EMSS (Gioia et al., 1990, Stocke et al., 1991). To this
sample we added 4 sources whose simultaneous spectra were obtained
by Stocke et
al. (1985). The aim was to study the spectral slope distribution at
radio frequencies without possible biases introduced by the
flat-radio-spectrum criterion often used in selecting BL Lac samples. 
We found
that about 15\% of the sources have steep spectra.  We considered 
the archived non-simultaneous data as well, finding an even higher
percentage of steep spectra.  This effect could be 
ascribed to variability, but other factors seem more
significant in determining the measured slope of spectra. The higher
percentage of steep spectra objects ($\sim$ 38\%), obtained by using
the 20cm NVSS observations, implies the presence of a large
fraction of extended flux that is preferentially detected by the VLA D
configuration. This seems also supported by a possible correlation
between spectral index and 20cm extended luminosities, computed either
with simultaneous or non-\-simultaneous data.

We find that the possible biases introduced by the flat--radio-spectrum
criterion in the RBLs samples cannot easily account for the discrepancies
observed in the $\langle V_e/V_a \rangle$ values of RBLs and XBLs samples
and therefore the issue of evolution of different BL Lac classes is still
open.
The percentages of steep spectrum BL Lacs found for the EMSS sample
cannot be applied straightforwardly to RBL samples, since a) the selection
by radio spectral index is sometimes computed using non-simultaneous data; b) the XBLs
properties are different from those of RBLs in the sense that RBLs are more 
variable and possibly more core dominated.
If the observed spread in $\alpha_r$ is mostly due to variability, we 
should expect a larger fraction of steep RBLs than XBLs.
If most of the effect is due, as the data seem to suggest, to the presence
of a significant extended component, that is lesser in RBL, the influence
on the radio selected objects should be less severe.

We stress here that the $\alpha_r$ distribution in the EMSS BL
Lacs simultaneous data is a continuous one and so any division in two
subsamples is somewhat arbitrary and does not have a physical
underlying support.  We are tempted to suggest a different cut-off
value of $\alpha_r=0.7$ (as done e.g. by Perlman et al. 1998 in DXRBS)
that would then exclude $\leq$ 10\% of all BL Lacs based upon our
findings here. But, because this would also include a very large
number of ``normal'' radio galaxies, some of which with optical
featureless spectra (see e.g. discussions in Rector \& Stocke, 2001,
Perlman et al. 1996), other information would have to be used to
define them as BL Lacs (e.g. variable optical polarization).

However, we have shown that the selection effects applied by using 
a cut-off value in the radio spectral index of $\alpha_r=0.5$ are not
so severe. The steeper BL Lacs that are excluded from this criterion 
seem to have the same overall properties of the flat ones, especially
for what concerns the cosmological distribution. 
We can therefore conclude that the discrepancies found by using different
samples cannot be ascribed to the flat radio spectrum criterion.

\begin{acknowledgements}
It is pleasure to thank Tommaso Maccacaro for stimulating discussions
and suggestions. We thank the referee for his comments that helped 
improve the paper.
This work has received partial financial support from the Italian
Space Agency (ASI) and MIUR.
\end{acknowledgements}

%
%

\end{document}